\documentclass[9pt,conference]{IEEEtran}
\IEEEoverridecommandlockouts

\usepackage{cite}
\usepackage{amsmath,amssymb,amsfonts}
\usepackage{algpseudocode}  
\usepackage{algorithm}
\usepackage{mathtools}      
\usepackage{amsthm}         
\usepackage{mathrsfs}       
\usepackage{bbm}            

\usepackage{graphicx}
\usepackage{textcomp}
\usepackage{xcolor}
\usepackage{booktabs}

\def\BibTeX{{\rm B\kern-.05em{\sc i\kern-.025em b}\kern-.08em
    T\kern-.1667em\lower.7ex\hbox{E}\kern-.125emX}}
\begin{document}

\title{Scalable Community Detection Using Quantum Hamiltonian Descent and QUBO Formulation}

\author{\IEEEauthorblockN{
Jinglei Cheng\IEEEauthorrefmark{4}\IEEEauthorrefmark{1},
Ruilin Zhou\IEEEauthorrefmark{2}\IEEEauthorrefmark{1}, 
Yuhang Gan\IEEEauthorrefmark{2}, 
Chen Qian\IEEEauthorrefmark{2},
Junyu Liu\IEEEauthorrefmark{4}}
\IEEEauthorblockA{\IEEEauthorrefmark{4}University of Pittsburgh \
\IEEEauthorrefmark{2}University of California, Santa Cruz\\
\IEEEauthorrefmark{1}Authors contributed equally to this research. Corresponding to: jic373@pitt.edu, junyuliu@pitt.edu
}
}

\maketitle

\begin{abstract}
We present a quantum-inspired algorithm that utilizes Quantum Hamiltonian Descent (QHD) for efficient community detection. 
Our approach reformulates the community detection task as a Quadratic Unconstrained Binary Optimization (QUBO) problem, and QHD is deployed to identify optimal community structures. 
We implement a multi-level algorithm that iteratively refines community assignments by alternating between QUBO problem setup and QHD-based optimization. 
Benchmarking shows our method achieves up to 5.49\% better modularity scores while requiring less computational time compared to classical optimization approaches. 
This work demonstrates the potential of hybrid quantum-inspired solutions for advancing community detection in large-scale graph data.
\end{abstract}


\section{Introduction}


Quantum computing with the principles of quantum mechanics can process information in different ways from classical computers. This is because qubits can exist in superposition, and therefore can represent multiple states simultaneously~\cite{nielsen_quantum_computation}. 
This capability, combined with entanglement and interference, allows quantum algorithms to solve certain classes of problems, such as integer factorization and combinatorial optimization, more efficiently than classical methods. 
Therefore, quantum computing has lead to advancements in cryptography~\cite{shor_algorithm}, optimization~\cite{farhi_qaoa}, machine learning~\cite{biamonte_quantum_machine_learning}, and materials science~\cite{kandala_materials_simulation}. 
In parallel, quantum-inspired algorithms bring similar quantum concepts to classical computing, with principles such as tunneling, adiabatic transitions, and Hamiltonian dynamics to address complex problems without requiring quantum hardware. 
These approaches have demonstrated enhanced performance in areas like traffic routing and combinatorial optimization, where classical systems emulate quantum behaviors to achieve good efficiency~\cite{tiwary_quantum_inspired_optimization}. 

\begin{figure}
    \centering
    \includegraphics[width=\linewidth]{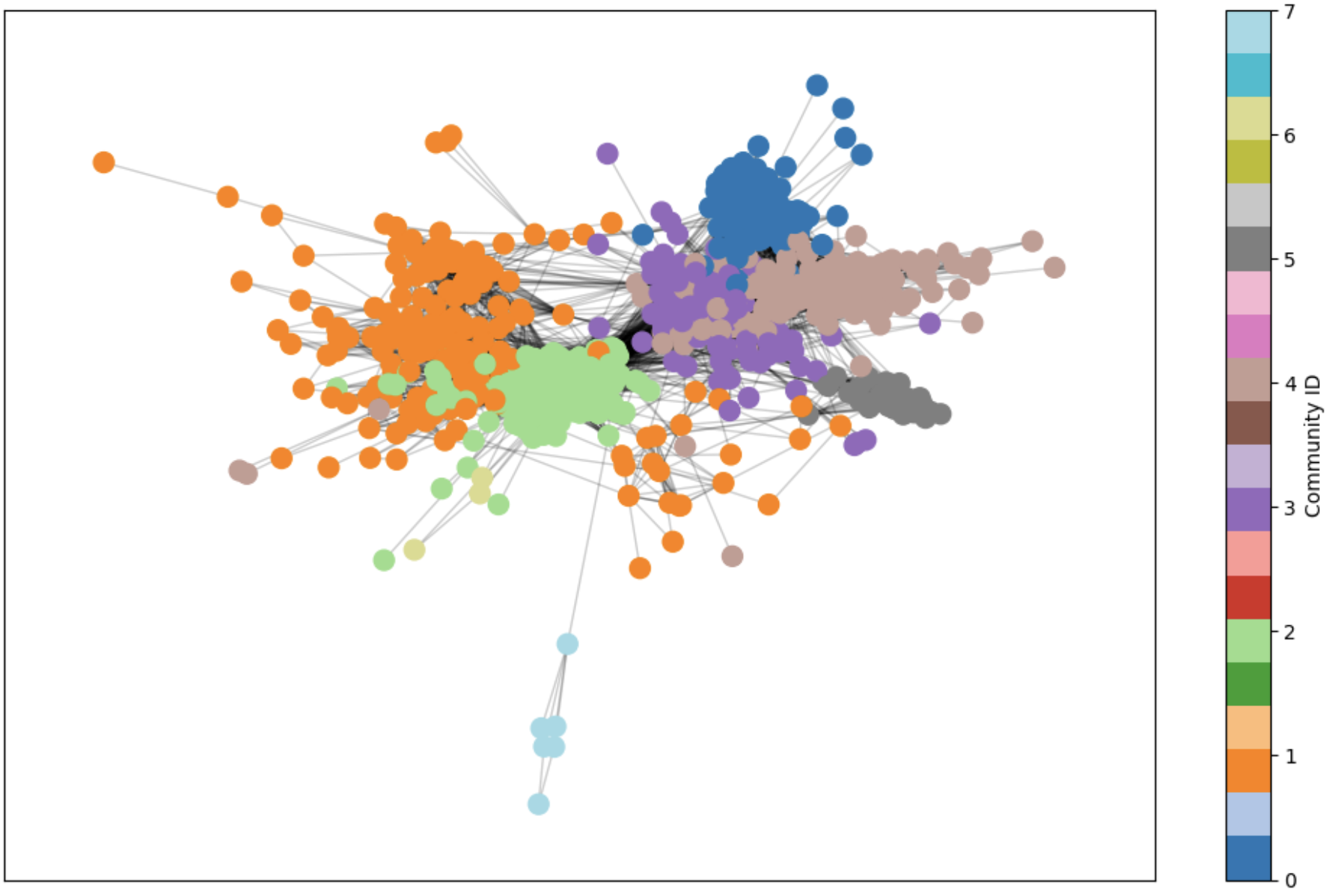}
    \caption{Visualization of community structure in a complex network. The network consists of nodes grouped into communities, each represented by a different color. }
    \label{fig:cd}
\end{figure}

Community detection is an important problem in network analysis, and it's essential for interpreting the structural and functional organization of complex systems~\cite{fortunato_community_detection}. 
An example is given in Figure~\ref{fig:cd}.
The overall goal is to identify communities—groups of nodes with denser connections among themselves than with the rest of the network~\cite{girvan_newman_modularity}. 
In large-scale graphs, which are common in various real-world applications such as social networks, biological systems, and communication networks, effective community detection provides valuable insights into network behavior and properties, and can support applications like anomaly detection, network optimization, and information dissemination~\cite{newman_networks}.
The impact of community detection is deep across many domains. In social networks, it enables the identification of groups with shared interests or behaviors, which can enhance recommendation systems and targeted marketing strategies~\cite{fortunato2016community, malliaros2013clustering}. In biological networks, community detection reveals functional modules or protein complexes, and contributes to a better understanding of cellular processes and disease mechanisms~\cite{ravasz2002hierarchical, barabasi2011network}. 
Additionally, fields like communication networks, financial systems, and transportation networks takes community detection to optimize efficiency, bolster security, and improve resilience against failures or attacks~\cite{newman2018networks, boccaletti2006complex}.


Detecting communities in large-scale graphs presents substantial computational challenges. Traditional algorithms often struggle with scalability, resulting in increased processing times and high resource consumption as network size grows~\cite{rossetti2018community}. Furthermore, maintaining the accuracy and quality of detected communities becomes more complex with larger datasets due to the intricate network structures and the presence of noise~\cite{fortunato2016community}. These challenges call for the development of more scalable algorithms capable of handling large-scale graphs without compromising performance or accuracy~\cite{abbe2017community}.


Quantum-inspired algorithms inherit conceptions from quantum computing to enhance classical computational methods~\cite{arrazola2019quantum,chepurko2022quantum,shaofaster,yelleti2023quantum,okawa2024quantum}. Quantum Hamiltonian Descent (QHD)~\cite{leng2023quantum,leng2024expanding} is an algorithm designed to efficiently solve optimization problems by leveraging quantum tunneling effects to escape local minima, and therefore enhances performance in non-convex optimization scenarios. By simulating the behavior of quantum systems, Quantum Hamiltonian Descent can navigate complex energy landscapes to find optimal solutions more effectively than some traditional optimization techniques~\cite{kushnir2024qhdopt}. This makes it a promising approach for tackling the computational demands of large-scale community detection, where the formulation is not always convex.

\begin{figure*}
    \centering
    \includegraphics[width=0.9\linewidth]{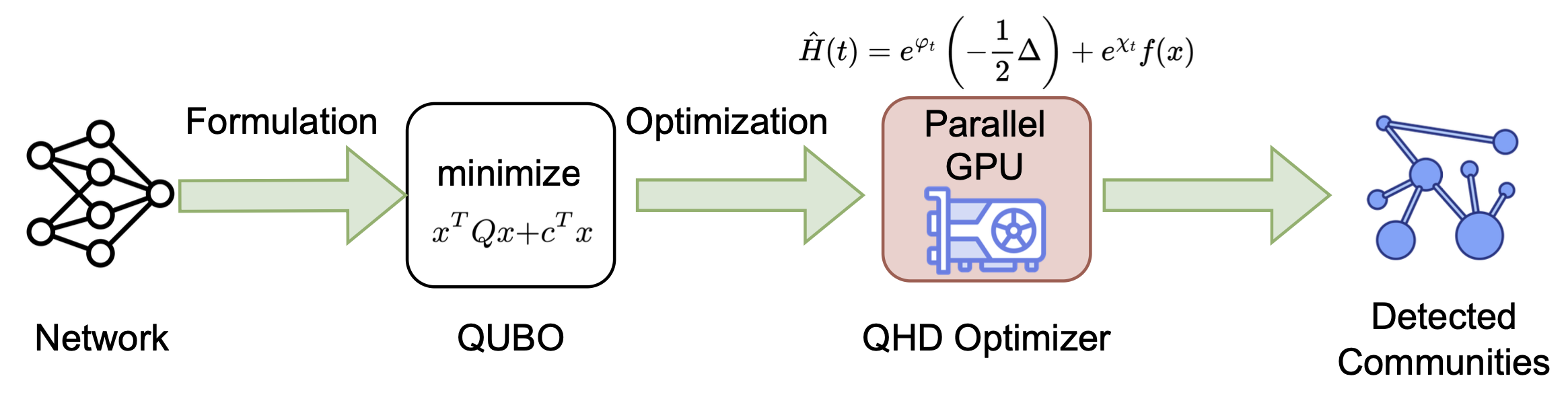}
    \caption{Overview of our quantum-inspired community detection approach. The input network is reformulated as a QUBO optimization problem, which is then solved using QHD accelerated by parallel GPU computation. The QHD optimizer uses quantum-inspired dynamics to efficiently identify lowest cost function values in the complex landscape, and delivers superior scalability for instances with thousands of nodes.}
    \label{fig:qhdcdteaser}
\end{figure*}

Our approach involves formulating the community detection problem as a QUBO model as shown in Figure~\ref{fig:qhdcdteaser}.
The QUBO formulation is a versatile framework for representing combinatorial optimization problems where the objective is to minimize a quadratic function of binary variables~\cite{kochenberger2014unconstrained}. 
By expressing community detection in this form, we can apply optimization algorithms like Quantum Hamiltonian Descent directly. 
In this way, the challenge of running graph algorithms on GPUs can be addressed by transforming them into a QUBO formulation, which allows for the deployment of GPU-accelerated algorithms.
Quantum Hamiltonian Descent is particularly effective for solving QUBO models due to its ability to efficiently explore huge and complex solution spaces. 
By mimicking the dynamics of quantum systems, it can avoid becoming trapped in local minima and converge rapidly to high-quality solutions. 
This efficiency is essential for large-scale problems where traditional optimization methods may be too slow or require excessive computational resources.


To further improve efficiency and solution quality, we have developed a multi-level algorithm that iteratively refines community assignments. This algorithm alternates between QUBO formulation and Quantum Hamiltonian Descent solving at different levels of graph granularity. Starting from a coarse representation of the graph, it progressively refines the community structure by incorporating more detailed information in each iteration. This hierarchical approach helps in escaping local optima and enhances the overall accuracy of the community detection process.

Our method has been benchmarked against GUROBI~\cite{GUROBI}, which is a leading commercial solver renowned for its optimization capabilities. 
On smaller problem instances, our algorithm matches GUROBI's performance, demonstrating its effectiveness and reliability. 
Notably, for larger problems exceeding 1,000 variables, our approach surpasses GUROBI.
This performance advantage is crucial for practical applications involving large-scale graphs where computational resources and time are significant constraints.
Recognizing the computational intensity of large-scale community detection, we have implemented our algorithm on multi-GPU architectures. 
GPUs offer massive parallel processing power, which accelerates the Quantum Hamiltonian Descent process and handles the extensive computations required for large graphs. Our implementation demonstrates excellent scalability and makes it feasible to analyze large networks in reasonable time budget. This efficient use of high-performance computing resources indicates the practicality of our approach for real-world applications.

Our contributions are:
\begin{itemize}
\item Developed a novel quantum-inspired community detection algorithm utilizing QHD optimization
\item Formulated community detection as a QUBO problem with an iterative multi-level refinement approach
\item Demonstrated competitive performance with GUROBI solver on small instances and superior scalability for problems exceeding 1,000 variables
\end{itemize}

The remainder of this paper is organized as follows. Section~\ref{background} provides the necessary background on community detection, QUBO problems, and quantum-inspired optimization techniques. 
In Section~\ref{formalization}, we present the formal mathematical framework of our approach, detailing the QUBO formulation. 
Section~\ref{implementation}  describes our implementation, including the multi-level algorithm design and QHD algorithm. 
We present comprehensive experimental results and performance analysis in Section~\ref{evaluation}. 
Finally, Section~\ref{conclusion} concludes the paper with a summary of our contributions and directions for future research.

\section{Background}
\label{background}
\subsection{Quantum Hamiltonian Descent}
QHD introduces a new quantum counterpart to classical gradient descent methods, derived through path integral quantization of the continuous-time limit of gradient descent algorithms. 
Unlike conventional approaches that only quantize specific components of classical algorithms, QHD quantizes the entire dynamical system, resulting in a quantum evolution governed by a Hamiltonian $\hat{H}(t) = e^{\phi_t}(-\frac{1}{2}\Delta) + e^{\chi_t}f(x)$, where $e^{\phi_t}$ and $e^{\chi_t}$ are damping parameters controlling system energy flow. 
This formulation allows QHD to take advantages of quantum tunneling effects to escape local minima by considering all possible paths at the same time, including those prohibited in classical mechanics. 
The algorithm exhibits three distinct phases - kinetic, global search, and descent - and has demonstrated superior performance over both classical gradient-based methods and quantum adiabatic algorithms in solving non-convex optimization problems. 
\subsection{Community Detection}
\label{background-cn}
Community Detection (CD), also known as graph clustering or graph partition, is one fundamental task in graph theory and network science with the main goal of identifying groups of nodes within a graph that are more densely connected to each other than the rest. 
Applications of community detection span various fields, such as social network analysis~\cite{azaouzi2019community}, biology~\cite{khawaja2024exploring}, computer networks, and recommendation systems~\cite{gasparetti2020community}. Based on the definition of the nodes, applications, and the definition of a good partition. Various methods have been proposed, such as the modularity-based method~\cite{newman2004finding}, which aims to maximize modularity, which is a measure of the strength of the division of a network into communities, spectral clustering, which involves the use the eigenvalues and eigenvectors of matrices like the Laplacian derived from the graph to identify community structures based on the network’s spectral properties and hierarchical clustering which use methods such as divisive clustering to build a hierarchy of communities either by progressively merging nodes (bottom-up) or recursively splitting larger communities (top-down) based on similarity measures.

\section{Formalization}
\label{formalization}
In this section, we introduce how we transform the CD problem into QUBO form and the main considerations of our formulations. 
\subsection{Modularity}
Formally, the community detection problem with the goal of maximizing modularity can be formulated as follows: Given an undirected graph $G = (V, E)$ with $n = |V|$ vertices and $m = |E|$ edges, the task of community detection is to partition the vertices into $k$ non-empty communities such that nodes within the same community are more densely connected while nodes in different communities are sparsely connected. One commonly used metric to measure how well the graph is partitioned is 'modularity' which is defined by:  
\begin{equation}
Q = \frac{1}{2m} \sum_{i,j} \left(A_{ij} - \frac{d_i d_j}{2m}\right)\delta(c_i, c_j)
\label{eq:modularity}
\end{equation}
where $A_{ij}$ is the element of the adjacency matrix, $d_i$ and $d_j$ are the degrees of nodes $i$ and $j$. And $\delta(c_i c_j)$ is the Kronecker delta function that equals 1 if nodes $i$ and $j$ are in the same community and 0 otherwise. 

Various heuristics have been proposed to address the community detection problem. Cut-based methods aim to minimize inter-community edges while maintaining balanced partitions~\cite{shin2022graph}.  Modularity-based methods focus on maximizing the density of intra-community edges relative to a null model~\cite{rosvall2019different}.  Another approach involves using quantum annealing; however, this method introduces additional overhead due to the need to embed the problem instance onto the quantum annealer~\cite{fernandez2021community}. 

\subsection{QUBO Formulation for Community Detection }

Inspired by previous classical and quantum methods, we propose a quantum-inspired method that takes Quantum Hamiltonian Descent to replace the original compute-intensive components of classical algorithms. This method first transforms the CD problem into a QUBO formulation, which is the first step for the use of both classical and quantum-inspired solvers that can be executed on GPUs without the need for direct embedding onto quantum hardware. The specific formalization is as follows.

\begin{algorithm}
\label{alg:1}
\caption{QUBO Construction for Community Detection}
\begin{algorithmic}[1]
\Require
    \State $n$: number of nodes
    \State $k$: number of communities
    \State $B$: modularity matrix
    \State $G(V,E)$: input graph
    \State $w_1, w_2, w_3$: penalty weights
\Ensure QUBO matrix $Q$ and linear terms $b$

\State Initialize $Q \in \mathbb{R}^{nk \times nk}, b \in \mathbb{R}^{nk}$ to zero

\Function{idx}{$i,c$}
    \Return $i \cdot k + c$
\EndFunction

\State // Modularity term
\State $Q_{\text{idx}(i,c),\text{idx}(j,c)} \gets -w_1B_{ij}$

\State // Assignment constraints
\State $Q_{\text{idx}(i,c_1),\text{idx}(i,c_2)} \gets 2w_2$
\State $b_{\text{idx}(i,c)} \gets -w_2$

\State // Cut penalty
\State $i_c \gets \text{idx}(u,c), j_c \gets \text{idx}(v,c)$
\State $Q_{i_c,j_c} \gets -2w_3$

\end{algorithmic}
\end{algorithm}

\subsubsection{Direct QUBO Formulation for Small Networks}
To adapt the modularity optimization problem for QUBO solvers, we construct a QUBO objective function that encapsulates both the modularity measure and the necessary constraints for valid community assignments. Let  binary variables $x_{i,c} \in \{0,1\}$, where:
\[
x_{i,c} = \begin{cases}
1 & \text{if vertex } i \text{ is assigned to community } c, \\
0 & \text{otherwise}.
\end{cases}
\]
Assume a maximum of $k$ communities, with $c \in \{1, 2, \dots, k\}$.
The modularity contribution can be expressed using the binary variables as:
\begin{equation}
Q_M = \frac{1}{2m} \sum_{i,j \in V} \left(A_{ij} - \frac{d_i d_j}{2m}\right) \sum_{c=1}^{k} x_{i,c} x_{j,c}.
\label{eq:modularity_qubo}
\end{equation}. Different from the previous formalization, which only focused on maximizing modularity, we also added other terms. To ensure that each vertex is assigned to exactly one community, we impose the following constraint:
\begin{equation}
Q_A = \lambda_A \sum_{i \in V} \left(1 - \sum_{c=1}^{k} x_{i,c}\right)^2,
\label{eq:assignment_constraint}
\end{equation}
where $\lambda_A$ is a penalty coefficient that enforces the constraints, and expanding the squared term ensures that derivation from the constraint incur a quadratic penalty. 
To prevent trivial solution where all vertices are assigned to a single community or most vertices are assigned to several communities and aim to provide a balance size of communities, we also introduce the constraints on the size of communities: 
\begin{equation}
Q_S = \lambda_S \sum_{c=1}^{k} \left(\sum_{i \in V} x_{i,c} - \frac{n}{k}\right)^2,
\label{eq:size_constraint}
\end{equation}
where $ \lambda_S$ is the penalty coefficient for balancing community sizes and $\frac{n}{k}$ represents the desired average community size. 
Based on these, the complete QUBO objective function to be maximized can be formed as 
\begin{equation}
Q = -Q_M + Q_A + Q_S,
\label{eq:complete_qubo}
\end{equation}
where $-Q_M$ corresponds to the maximization of the modularity, $Q_A$ enforces the constraint that each vertex will be in exactly one community, and $Q_S$ enforces a balanced constraint, which is also a consideration in most CD formalization. In general, our formalization gives comprehensive objective functions that take multiple aspects into consideration instead of solely maximizing the modularity metric. More details are shown in Algorithm~\ref{alg:1}.

\subsubsection{Scale to Larger Networks}

\begin{algorithm}
\label{algo-2}
\caption{Multilevel Community Detection}
\begin{algorithmic}[1]
\Require
\State $G_0(V_0,E_0)$: input graph
\State $k$: number of communities
\State $\theta$: coarsening threshold
\Ensure Community assignments $P: V_0 \rightarrow {1,\dots,k}$
\Function{Coarsen}{$G_i$}
\State Match nodes in $G_i$ to form super-nodes
\State Combine matched nodes to create $G_{i+1}$
\Return $G_{i+1}$
\EndFunction
\Function{Project}{$P_{i+1}, G_i$}
\State Map community assignments from level $i+1$ to $i$
\Return $P_i$
\EndFunction
\State // Coarsening Phase
\State $i \gets 0$
\While{$|V_i| > \theta$}
\State $G_{i+1} \gets $ \Call{Coarsen}{$G_i$}
\State $i \gets i + 1$
\EndWhile
\State $m \gets i$ // Final level
\State // Initial Solution
\State $P_m \gets $ \Call{SolveBase}{$G_m, k$}
\State // Uncoarsening Phase
\For{$i = m-1$ \textbf{downto} $0$}
\State $P_i \gets $ \Call{Project}{$P_{i+1}, G_i$}
\State $P_i \gets $ \Call{Refine}{$P_i, G_i, k$}
\EndFor
\State \Return $P_0$
\end{algorithmic}
\end{algorithm}

While previous direct QUBO formalization is effective for small to medium-sized networks($|V| \leq 1000$), larger networks require more scalable approaches. To address this and inspired by classical methods~\cite{metis}, we employ a hierarchical methodology that maintains solution quality while reducing computational complexity. More specifically, our method includes the following steps: Coarsening, initial partition, uncoarsening, and refinement. More details are shown in Algorithm~\ref{algo-2}.

\textbf{Coarsening Phase}: We iteratively reduce the graph size by aggregating vertices into super-nodes, thereby preserving the community structure at multiple scales. At each coarsening step, a heavy-edge matching strategy is utilized to select vertex pairs with strong connectivity:
\begin{equation}
w(e) = \alpha \frac{|N(u) \cap N(v)|}{|N(u) \cup N(v)|} + \beta \frac{A_{uv}}{\max_{e \in E} A_{uv}},
\label{eq:matching_weight}
\end{equation}
where $N(u)$ denotes the neighborhood of vertex $u$.$\alpha$, and $\beta$ are weighting coefficients that balance the contribution of neighborhood overlap and edge weight. This coarsening phase reduces the number of vertices while preserving the essential community structure. 

\textbf{Initial Partition}: Once the graph is sufficiently coarsened to a manageable size,  we apply the direct QUBO formulation (\ref{eq:complete_qubo}) to partition the coarsest graph. This step provides an initial community assignment that captures the high-level structure of the original graph.

\textbf{Uncoarsening and Refinement}: After initial partitioning is found in the coarsest graph, the partitioning is then projected back to the original graph through the following steps:
\begin{enumerate}
    \item \textbf{Projection}: Starting from the coarsest level, iteratively map the community from each coarsened graph  \( G_i \) to the next finer graph \( G_{i-1} \). This is based on the super-node correspondence established during the coarsening phase which can effectively transfer the high-level community structure to more detailed graph levels. 
    \item \textbf{Refinement}: At each level, we optimize the community assignments by iteratively reassigning nodes to communities that result in the highest gain in modularity. This involves evaluating potential moves for each node and updating its community assignment if the move improves the overall modularity. The refinement process continues until certain iterations of the threshold are met or no further improvement is achieved. 
\end{enumerate}

\section{Implementation}
\label{implementation}

\subsection{QHD on QUBO}
QHDOPT~\cite{kushnir2024qhdopt} implements Quantum Hamiltonian Descent for solving nonlinear optimization problems. Its key innovation is implementing this optimization via quantum evolution governed by the time-dependent Schrödinger equation:
$$
i\frac{\partial}{\partial t}\Psi(t,x) = \left(e^{\phi_t}\left(-\frac{1}{2}\Delta\right) + e^{\chi_t}f(x)\right)\Psi(t,x)
$$
where $\Delta$ is the Laplacian operator and $e^{\phi_t}, e^{\chi_t}$ control the energy distribution of the quantum system.

A significant computational advantage of this approach is that it requires only matrix multiplication operations, avoiding the need to solve linear systems (Ax=b). This makes the method particularly amenable to modern hardware acceleration. For practical implementation, QHDOPT employs a discretization and embedding strategy where the continuous Hamiltonian is discretized into an $N^n$-dimensional operator:
$$
\hat{H}(t) = e^{\phi_t}\left(-\frac{1}{2}L_d\right) + e^{\chi_t}F_d
$$
where $L_d$ represents the discretized Laplacian and $F_d$ encodes the objective function. 

The method's matrix-multiplication-based formulation opens up possibilities for various compression techniques that can transform large sparse problems into smaller, denser ones. This includes techniques like tensor network compression~\cite{cichocki2016tensor}, hierarchical matrix approximations~\cite{hackbusch2015hierarchical}, and randomized sketching methods~\cite{woodruff2014sketching}. Any additional constraints created during problem transformation can be efficiently handled through penalty-based methods, avoiding the need for complex constraint satisfaction algorithms.

When dealing with QUBO problems, QHDOPT can be efficiently accelerated using GPUs through careful implementation of the quantum Hamiltonian dynamics. The key is mapping the QUBO problem $\min_{x \in \{0,1\}^n} x^T Q x + b^T x$ to QHD's Hamiltonian framework in a way that exploits massive parallelism. Sparse operations can be particularly accelerated using specialized GPU packages like cuSPARSE~\cite{cuSPARSE} or MAGMA~\cite{magma}, which provide optimized implementations for sparse matrix operations. The time evolution can be implemented using GPU-optimized linear algebra operations, with the Laplacian operator $\Delta$ computed using parallel finite difference schemes and the potential term $e^{\chi_t}f(x)$ evaluated using batched matrix operations.

By leveraging frameworks like PyTorch~\cite{NEURIPS2019_9015} or JAX~\cite{jax2018github} that provide automatic differentiation and GPU acceleration, the quantum dynamics can be simulated efficiently - the gradient computations for the Hamiltonian evolution can be parallelized across the quantum state dimensions, and the time evolution steps can be batched for exploring multiple initial conditions simultaneously. The classical refinement step in QHDOPT can utilize GPU-accelerated optimizers to efficiently project solutions back to binary values. This hybrid quantum-classical approach combines quantum effects with massive parallel computing capabilities, making it particularly effective for large-scale optimization problems where traditional methods might struggle.

\begin{figure*}
\vspace{-3mm}

    \centering
    \includegraphics[width=0.8\textwidth]{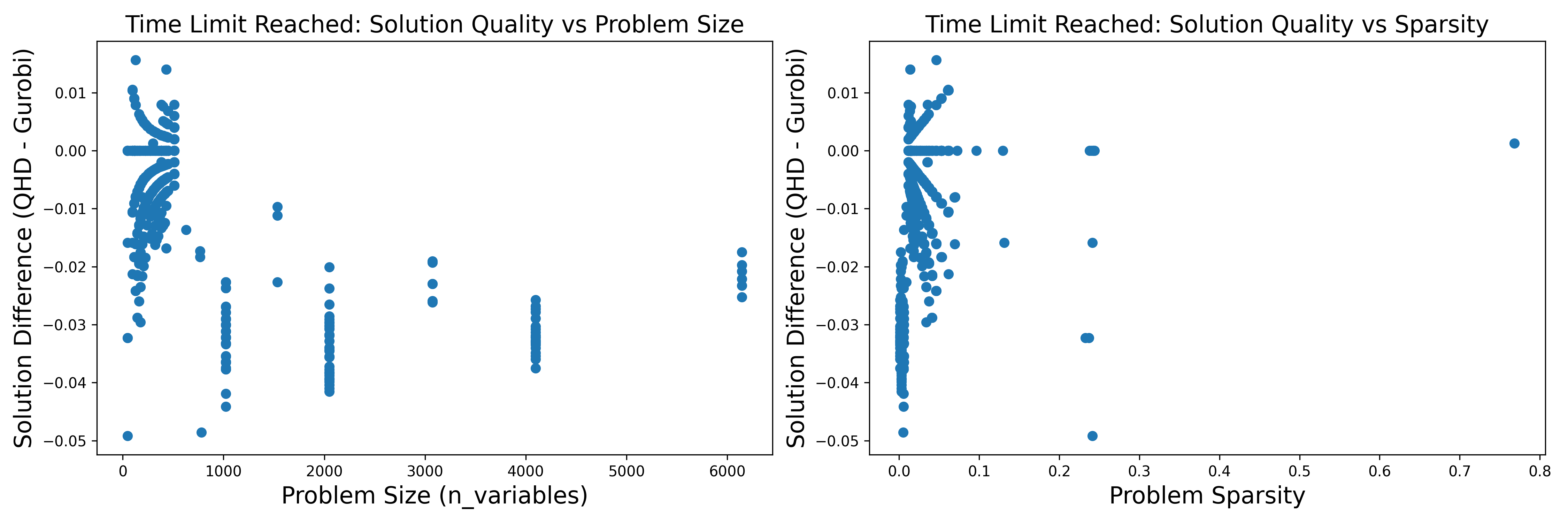}
    \caption{Solution Quality Comparison When GUROBI Hits Time Limit: For 739 instances where GUROBI exceeded time limits (predominantly in larger problems), QHD found better solutions in 71.4\% of cases. For larger-scale problems, QHD demonstrated superior performance by finding better solutions than GUROBI within the same time constraints. This advantage stems from QHD's inherent parallel processing capabilities, which can be further enhanced through additional GPU resources and optimized sparse matrix operations, suggesting even greater potential for scaling to larger problem instances.}
    \label{fig:timelimit}
\end{figure*}

\subsection{Multi-level Community Detection}

This multilevel community detection algorithm is designed to efficiently handle large-scale graphs by employing a hierarchical approach. The algorithm operates in three main phases: First, in the coarsening phase, it repeatedly combines nodes to create a hierarchy of progressively smaller graphs while preserving the community structure. Once the graph is sufficiently small, it solves the community detection problem directly on this coarsened graph. Finally, in the uncoarsening phase, it progressively maps the solution back to finer levels while refining the communities at each step. This multilevel approach allows the algorithm to maintain both computational efficiency and solution quality by working with smaller problems during the initial solution while preserving the ability to make fine-grained adjustments during refinement.

\section{Evaluation}
\label{evaluation}

\subsection{Evaluation Setting}

The experimental evaluation was conducted on a system running Debian Linux with kernel version 5.10.0-22-amd64. The computing platform features a dual 16-core CPU (32 cores total). The server is equipped with 251 GB of RAM.
Implementation of QHD on QUBO is running with four NVIDIA A5000 GPUs and the comparison with GUROBI is conducted with CPUs as mentioned.

\subsection{QUBO Solver with GUROBI and QHD}

Comparing the computational performance between GUROBI and QHD presents unique challenges due to their different control parameters. 
While GUROBI can be configured with node exploration limits and termination times, QHD allows adjustment of sample sizes and iteration counts. 
To establish comparisons, we adopted a time-based benchmarking approach: first measuring QHD's execution time, then allocating GUROBI the same as its time limit. This methodology is justified because superior performance can be demonstrated either through faster execution at equal solution quality or through better solutions within equal time constraints. Our experiments focused on the latter approach, comparing solution quality between the two solvers under equivalent time constraints. The results demonstrate that QHD achieves higher solution quality while GUROBI cannot finish with status as OPTIMAL within the allocated time. It's worth noting that QHD's performance can be further enhanced with additional computational resources due to its inherent massive parallelism capabilities.
The results in Figure~\ref{fig:Optimal Solutions_analysis} and Figure~\ref{fig:timelimit} demonstrate robust performance patterns across a substantial dataset of 938 instances. In the 199 cases where GUROBI found proven optimal solutions (typically smaller problems with mean size of 54 variables), QHD matched these optimal solutions in 75.4\% of cases. More importantly, in the larger set of 739 instances where GUROBI hit its time limit (predominantly larger problems with mean size of 614 variables), QHD demonstrated superior performance by finding better solutions in 71.4\% of cases and matching GUROBI's solutions in another 17.2\%. While the time-limited instances show lower sparsity (mean 0.028 vs 0.157), this is primarily a characteristic of larger problem instances rather than a direct driver of performance difference. The results, based on this comprehensive dataset, strongly indicate QHD's effectiveness for larger-scale optimization problems where traditional solvers face computational limitations, regardless of sparsity.

\subsection{Evaluation on Small Size Networks}

\begin{figure*}
\vspace{-3mm}
    \centering
    \includegraphics[width=0.7\textwidth]{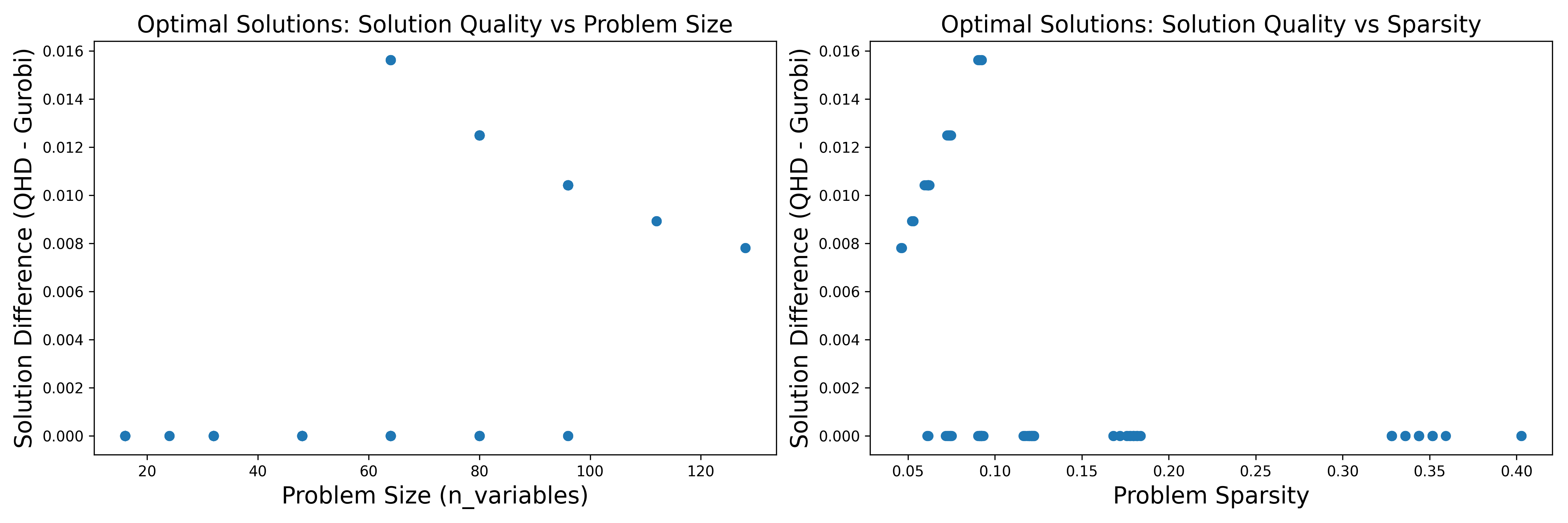}
    \caption{Solution Quality Comparison When GUROBI Reaches Optimality: Among 199 instances where GUROBI found optimal solutions (predominantly in smaller problems), QHD achieved identical solutions in 75.4\% of cases. In cases where QHD did not match GUROBI's optimal solutions (24.6\% of optimally solved instances), the objective value differences remained within a 1.6\% relative gap, primarily occurring in smaller instances where the absolute differences in objective values were minimal.
}
    \label{fig:optimal}
\end{figure*}

\begin{figure}[htb]
\vspace{-3mm}

    \centering
    \includegraphics[width=0.8\linewidth]{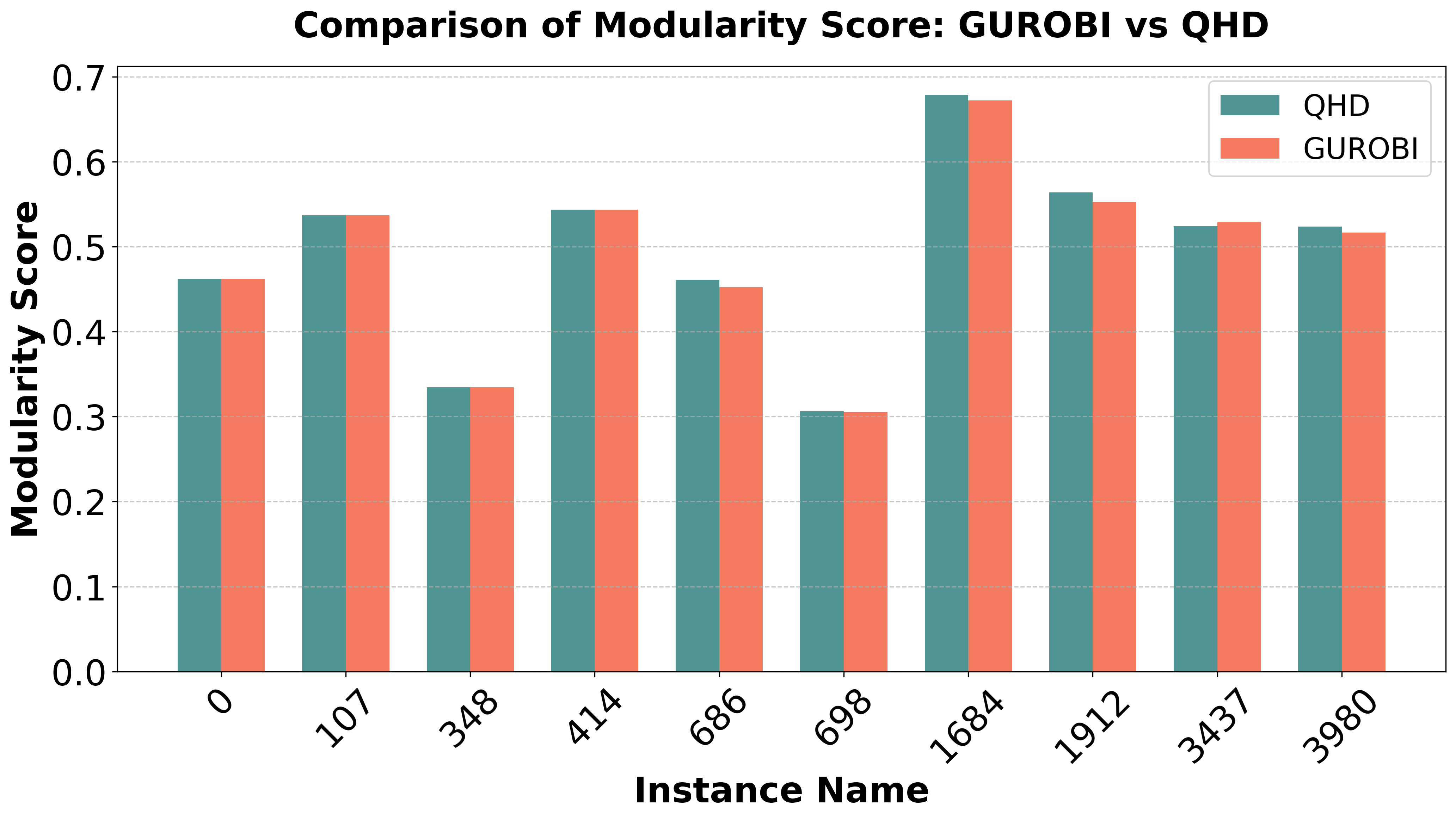}
    \caption{Performance comparison between QHD and GUROBI on network instances ranging from 52 to 1,034 nodes. QHD achieves superior modularity scores in 80\% of test cases with an average improvement of 0.0029, while requiring only 20\% of GUROBI's computational time using four GPUs. Results demonstrate QHD's effectiveness across different network densities (3.4\%-15.2\%).}
    \label{fig:modularity}
\end{figure}

Our experimental evaluation compared the Quantum Hamiltonian Descent approach with the GUROBI-based modularity optimization across a diverse set of network instances. As shown in Table~\ref{tab:small_results}, the test bed has 10 networks ranging from a graph with (52 nodes and 146 edges) to an instance of (1,034 nodes, 26,749 edges), with edge densities from 3.4\% to 15.2\%. As shown in Fig.~\ref{fig:modularity}, the QHD-based method demonstrated robust performance, achieving higher modularity scores in 8 out of 10 instances compared to the GUROBI implementation. The average modularity difference of 0.0029 in favor of QHD suggests that quantum-inspired optimization can effectively match or slightly exceed traditional exact optimization approaches. This robust performance, where QHD achieves higher modularity scores in 8 out of 10 test instances while using only 20\% of GUROBI's computational time with four GPUs, demonstrates its practical advantages over classical modularity optimization methods, with the potential for even greater efficiency gains through increased GPU parallelization.

\begin{table}[h]
\vspace{-1mm}
\caption{Instance Properties and Modularity Scores}
\centering
\begin{tabular}{cccccc}

\toprule
Instance & Nodes & Edges & Density \% & GUROBI & QHD \\
\midrule
0 & 333 & 2,519 & 4.56 & 0.4523 & \textbf{0.4610} \\
107 & 1,034 & 26,749 & 5.01 & \textbf{0.5290} & 0.5241 \\
348 & 224 & 3,192 & 12.78 & 0.3055 & \textbf{0.3063} \\
414 & 150 & 1,693 & 15.15 & 0.5438 & \textbf{0.5438} \\
686 & 168 & 1,656 & 11.80 & 0.3347 & \textbf{0.3347} \\
698 & 61 & 270 & 14.75 & 0.5369 & \textbf{0.5369} \\
1684 & 786 & 14,024 & 4.55 & 0.5528 & \textbf{0.5640} \\
1912 & 747 & 30,025 & 10.78 & 0.5167 & \textbf{0.5239} \\
3437 & 534 & 4,813 & 3.38 & 0.6724 & \textbf{0.6784} \\
3980 & 52 & 146 & 11.01 & \textbf{0.4619} & 0.4619 \\
\bottomrule
\end{tabular}
\label{tab:small_results}
\end{table}

\subsection{Evaluation on Large Size Networks}

\begin{figure}[htb]
\vspace{-3mm}

    \centering
    \includegraphics[width=0.8\linewidth]{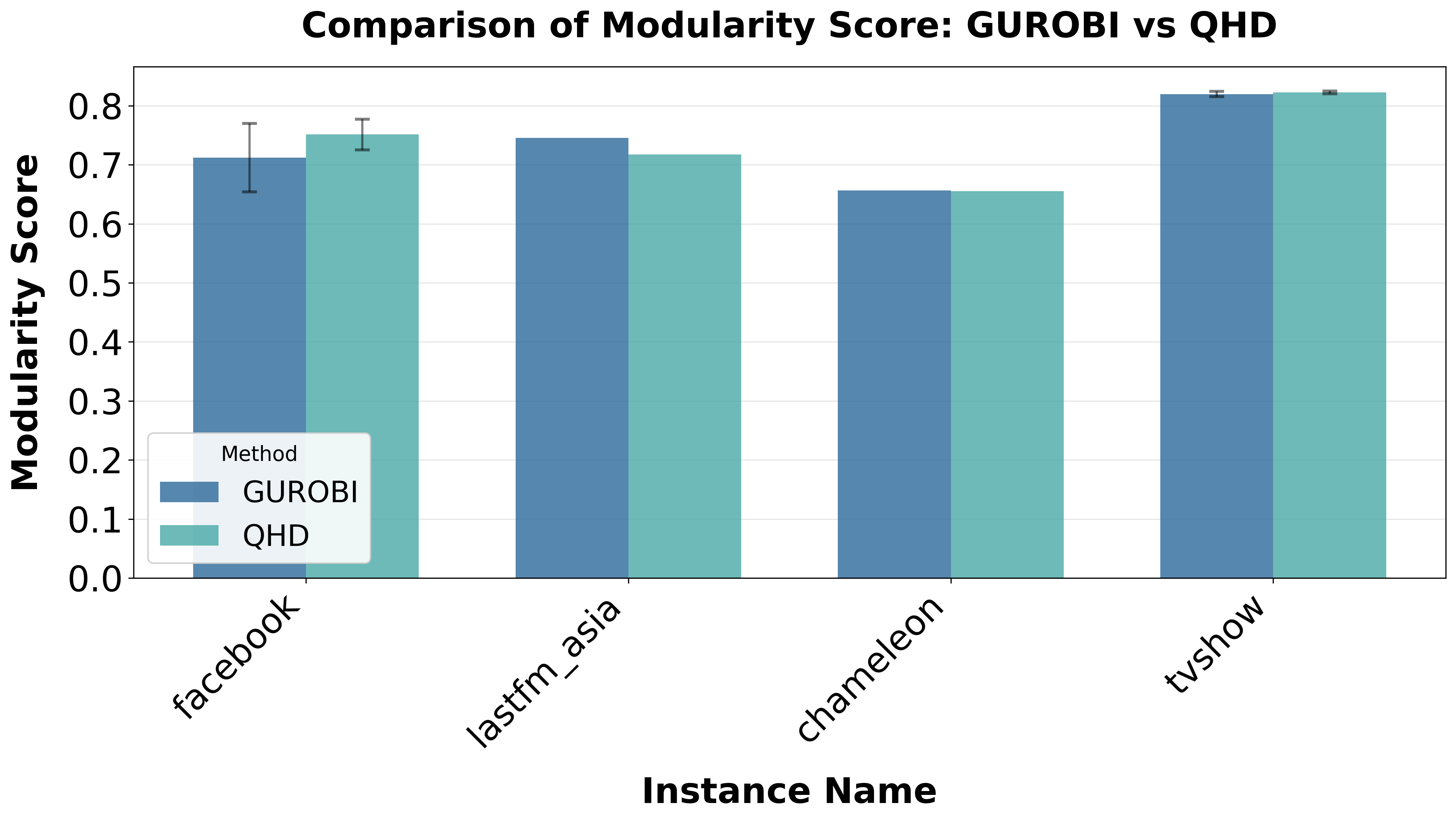}
    \caption{The performance advantage varies with network density, from QHD's 5.49\% improvement on Facebook (density 0.0108) to GUROBI's 3.79\% advantage on sparse LastFM (density 0.0010). Both methods show comparable performance on medium-density networks.}
    \label{fig:large}
\end{figure}

Table~\ref{tab:large} and Figure~\ref{fig:large} of GUROBI and QHD performance across different networks reveal a significant density-dependent pattern. In the Facebook network (density 0.0108), QHD achieves a notable 5.49\% higher modularity score (0.7512 ± 0.0258) compared to GUROBI (0.7121 ± 0.0579), while also demonstrating better stability with lower variance. The densest network, Chameleon (density 0.0121), shows nearly identical performance between the methods, with only a 0.19\% difference favoring GUROBI. The TVShow network, despite its lower density (0.0023), exhibits near performance from both methods, with QHD showing a marginal advantage of 0.33\% (QHD: 0.8223 ± 0.0025, GUROBI: 0.8196 ± 0.0044) and both methods maintaining very low variance. Notably, in the sparsest network, LastFM (density 0.0010), GUROBI outperforms QHD by 3.79\%. These results indicate that QHD's advantages are most pronounced in moderately dense networks. And GUROBI maintains good performance across network types, particularly in sparser networks, which is attributed to its branch-and-cut algorithm. 

\begin{table}[htb]
\vspace{-1mm}
\centering
\caption{Comparison of Graph Properties and Modularity Scores}
\begin{tabular}{cccccc}
\toprule
Instance & Nodes & Edges & Density \% & GUROBI & QHD \\
\midrule
facebook & 4,039 & 88,234 & 1.08 & 0.7121 & \textbf{0.7512} \\
lastfm\_asia & 7,626 & 27,807 & 0.10 & \textbf{0.7455} & 0.7172 \\
musae\_chameleon & 2,279 & 31,372 & 1.21 & \textbf{0.6567} & 0.6554 \\
tvshow & 3,894 & 17,240 & 0.23 & 0.8196 & \textbf{0.8223} \\
\bottomrule
\end{tabular}

\label{tab:large}
\end{table}

\vspace{-1mm}
\section{Conclusion}
\label{conclusion}

This work demonstrates the effectiveness of quantum-inspired optimization for community detection through our QHD-based approach. By reformulating community detection as a QUBO problem and leveraging GPU-accelerated quantum-inspired optimization, we achieve comparable or superior performance to traditional methods while reducing computational time. Our experimental results show strong performance on moderately dense networks, with up to 5.49\% improvement in modularity scores and enhanced stability compared to GUROBI. The implementation of multi-GPU acceleration suggests further scalability potential for larger networks. These findings highlight the promising intersection of quantum-inspired algorithms and high-performance computing for complex network analysis tasks. Future work could explore extending this approach to other graph optimization problems and investigating additional acceleration techniques for handling ultra-large-scale networks. Better designed algorithms to formulate graphs into denser matrices can reduce the number of variables in QUBO, and combination with high-performance sparsity computation will also be helpful. Our results demonstrate that quantum-inspired methods, combined with modern computing architectures, offer a practical pathway for advancing the capabilities of community detection in real-world network analysis applications.

\vspace{-1mm}
\section{Ackowledgement}
JL is supported in part by the University of Pittsburgh, School of Computing and Information, Department of Computer Science, Pitt Cyber, PQI Community Collaboration Awards. And by NASA under award number 80NSSC25M7057. This research used resources of the Oak Ridge Leadership Computing Facility, which is a DOE Office of Science User Facility supported under Contract DE-AC05-00OR22725. RL is supported by NSF Grants 2322919, 2420632, 2426031, 2426940, 2114113, and DOE Grant DE-SC0022069. We thank Yuxiang Peng from University of Maryland and Jiaqi Leng from UC Berkely for their helpful and insightul discussions.

\bibliographystyle{IEEEtran}  
\bibliography{ref}         

\begin{thebibliography}{10}
\providecommand{\url}[1]{#1}
\csname url@samestyle\endcsname
\providecommand{\newblock}{\relax}
\providecommand{\bibinfo}[2]{#2}
\providecommand{\BIBentrySTDinterwordspacing}{\spaceskip=0pt\relax}
\providecommand{\BIBentryALTinterwordstretchfactor}{4}
\providecommand{\BIBentryALTinterwordspacing}{\spaceskip=\fontdimen2\font plus
\BIBentryALTinterwordstretchfactor\fontdimen3\font minus \fontdimen4\font\relax}
\providecommand{\BIBforeignlanguage}[2]{{%
\expandafter\ifx\csname l@#1\endcsname\relax
\typeout{** WARNING: IEEEtran.bst: No hyphenation pattern has been}%
\typeout{** loaded for the language `#1'. Using the pattern for}%
\typeout{** the default language instead.}%
\else
\language=\csname l@#1\endcsname
\fi
#2}}
\providecommand{\BIBdecl}{\relax}
\BIBdecl

\bibitem{nielsen_quantum_computation}
M.~A. Nielsen and I.~L. Chuang, \emph{Quantum Computation and Quantum Information}, 10th~ed.\hskip 1em plus 0.5em minus 0.4em\relax Cambridge, UK: Cambridge University Press, 2010.

\bibitem{shor_algorithm}
P.~W. Shor, ``Algorithms for quantum computation: Discrete logarithms and factoring,'' in \emph{Proceedings of the 35th Annual Symposium on Foundations of Computer Science}.\hskip 1em plus 0.5em minus 0.4em\relax IEEE, 1994, pp. 124--134.

\bibitem{farhi_qaoa}
\BIBentryALTinterwordspacing
E.~Farhi, J.~Goldstone, and S.~Gutmann, ``A quantum approximate optimization algorithm,'' \emph{arXiv preprint arXiv:1411.4028}, 2014. [Online]. Available: \url{https://arxiv.org/abs/1411.4028}
\BIBentrySTDinterwordspacing

\bibitem{biamonte_quantum_machine_learning}
J.~Biamonte, P.~Wittek, N.~Pancotti, P.~Rebentrost, N.~Wiebe, and S.~Lloyd, ``Quantum machine learning,'' \emph{Nature}, vol. 549, pp. 195--202, 2017.

\bibitem{kandala_materials_simulation}
A.~Kandala, A.~Mezzacapo, K.~Temme, M.~Takita, M.~Brink, J.~M. Chow, and J.~M. Gambetta, ``Hardware-efficient variational quantum eigensolver for small molecules and quantum magnets,'' \emph{Nature}, vol. 549, pp. 242--246, 2017.

\bibitem{tiwary_quantum_inspired_optimization}
A.~Laio and M.~Parrinello, ``Escaping free-energy minima,'' \emph{Proceedings of the national academy of sciences}, vol.~99, no.~20, pp. 12\,562--12\,566, 2002.

\bibitem{fortunato_community_detection}
S.~Fortunato, ``Community detection in graphs,'' \emph{Physics Reports}, vol. 486, no. 3--5, pp. 75--174, 2010.

\bibitem{girvan_newman_modularity}
M.~Girvan and M.~E.~J. Newman, ``Community structure in social and biological networks,'' \emph{Proceedings of the National Academy of Sciences}, vol.~99, no.~12, pp. 7821--7826, 2002.

\bibitem{newman_networks}
M.~E.~J. Newman, \emph{Networks: An Introduction}.\hskip 1em plus 0.5em minus 0.4em\relax Oxford, UK: Oxford University Press, 2010.

\bibitem{fortunato2016community}
S.~Fortunato and D.~Hric, ``Community detection in networks: A user guide,'' \emph{Physics Reports}, vol. 659, pp. 1--44, 2016.

\bibitem{malliaros2013clustering}
F.~D. Malliaros and M.~Vazirgiannis, ``Clustering and community detection in directed networks: A survey,'' \emph{Physics Reports}, vol. 533, no.~4, pp. 95--142, 2013.

\bibitem{ravasz2002hierarchical}
E.~Ravasz, A.~L. Somera, D.~A. Mongru, Z.~N. Oltvai, and A.-L. Barabasi, ``Hierarchical organization of modularity in metabolic networks,'' \emph{Science}, vol. 297, no. 5586, pp. 1551--1555, 2002.

\bibitem{barabasi2011network}
A.-L. Barabasi, N.~Gulbahce, and J.~Loscalzo, ``Network medicine: A network-based approach to human disease,'' \emph{Nature Reviews Genetics}, vol.~12, no.~1, pp. 56--68, 2011.

\bibitem{newman2018networks}
M.~Newman, \emph{Networks}.\hskip 1em plus 0.5em minus 0.4em\relax Oxford, UK: Oxford University Press, 2018.

\bibitem{boccaletti2006complex}
S.~Boccaletti, V.~Latora, Y.~Moreno, M.~Chavez, and D.-U. Hwang, ``Complex networks: Structure and dynamics,'' \emph{Physics Reports}, vol. 424, no. 4--5, pp. 175--308, 2006.

\bibitem{rossetti2018community}
G.~Rossetti and R.~Cazabet, ``Community discovery in dynamic networks: A survey,'' \emph{ACM Computing Surveys}, vol.~51, no.~2, pp. 1--37, 2018.

\bibitem{abbe2017community}
E.~Abbe, ``Community detection and stochastic block models: Recent developments,'' \emph{Journal of Machine Learning Research}, vol.~18, no.~1, pp. 6446--6531, 2017.

\bibitem{arrazola2019quantum}
J.~M. Arrazola, A.~Delgado, B.~R. Bardhan, and S.~Lloyd, ``Quantum-inspired algorithms in practice,'' \emph{Quantum}, vol.~4, p. 307, 2020.

\bibitem{chepurko2022quantum}
N.~Chepurko, K.~Clarkson, L.~Horesh, H.~Lin, and D.~Woodruff, ``Quantum-inspired algorithms from randomized numerical linear algebra,'' \emph{Proceedings of the 39th International Conference on Machine Learning}, vol. 162, pp. 3879--3900, 2022.

\bibitem{shaofaster}
C.~Shao and A.~Montanaro, ``Faster quantum-inspired algorithms for solving linear systems,'' \emph{arXiv preprint arXiv:2103.10309}, 2021.

\bibitem{yelleti2023quantum}
V.~Yelleti and R.~Krishna, ``Quantum-inspired evolutionary algorithms for feature subset selection: A comprehensive survey,'' \emph{arXiv preprint arXiv:2407.17946}, 2023.

\bibitem{okawa2024quantum}
H.~Okawa, Q.-G. Zeng, X.-Z. Tao, and M.-H. Yung, ``Quantum-annealing-inspired algorithms for track reconstruction at high-energy colliders,'' \emph{arXiv preprint arXiv:2402.14718}, 2024.

\bibitem{leng2023quantum}
J.~Leng, E.~Hickman, J.~Li, and X.~Wu, ``Quantum hamiltonian descent,'' \emph{arXiv preprint arXiv:2303.01471}, 2023.

\bibitem{leng2024expanding}
J.~Leng, J.~Li, Y.~Peng, and X.~Wu, ``Expanding hardware-efficiently manipulable hilbert space via hamiltonian embedding,'' \emph{arXiv preprint arXiv:2401.08550}, 2024.

\bibitem{kushnir2024qhdopt}
S.~Kushnir, J.~Leng, Y.~Peng, L.~Fan, and X.~Wu, ``Qhdopt: A software for nonlinear optimization with quantum hamiltonian descent,'' \emph{arXiv preprint arXiv:2409.03121}, 2024.

\bibitem{kochenberger2014unconstrained}
G.~Kochenberger, J.-K. Hao, F.~Glover, M.~Lewis, Z.~L{\"u}, and H.~Wang, ``The unconstrained binary quadratic programming problem: A survey,'' \emph{Journal of Combinatorial Optimization}, vol.~28, no.~1, pp. 58--81, 2014.

\bibitem{GUROBI}
\BIBentryALTinterwordspacing
{Gurobi Optimization, LLC}, ``{Gurobi Optimizer Reference Manual},'' 2024. [Online]. Available: \url{https://www.gurobi.com}
\BIBentrySTDinterwordspacing

\bibitem{azaouzi2019community}
M.~Azaouzi, D.~Rhouma, and L.~Ben~Romdhane, ``Community detection in large-scale social networks: state-of-the-art and future directions,'' \emph{Social Network Analysis and Mining}, vol.~9, no.~1, p.~23, 2019.

\bibitem{khawaja2024exploring}
F.~R. Khawaja, Z.~Zhang, Y.~Memon, and A.~Ullah, ``Exploring community detection methods and their diverse applications in complex networks: a comprehensive review,'' \emph{Social Network Analysis and Mining}, vol.~14, no.~1, p. 115, 2024.

\bibitem{gasparetti2020community}
F.~Gasparetti, G.~Sansonetti, and A.~Micarelli, ``Community detection in social recommender systems: a survey,'' \emph{Applied Intelligence}, vol.~51, pp. 3975--3995, 2021.

\bibitem{newman2004finding}
M.~E.~J. Newman and M.~Girvan, ``Finding and evaluating community structure in networks,'' \emph{Physical Review E}, vol.~69, no.~2, p. 026113, 2004.

\bibitem{shin2022graph}
H.~Shin, J.~Park, and D.~Kang, ``A graph-cut-based approach to community detection in networks,'' \emph{Applied Sciences}, vol.~12, no.~12, p. 6218, 2022.

\bibitem{rosvall2019different}
M.~Rosvall, J.-C. Delvenne, M.~T. Schaub, and R.~Lambiotte, ``Different approaches to community detection,'' \emph{Advances in network clustering and blockmodeling}, pp. 105--119, 2019.

\bibitem{fernandez2021community}
M.~Fern{\'a}ndez-Campoamor, C.~O'Meara, G.~Cortiana, V.~Peric, and J.~Bernab{\'e}-Moreno, ``Community detection in electrical grids using quantum annealing,'' \emph{arXiv preprint arXiv:2112.08300}, 2021.

\bibitem{metis}
\BIBentryALTinterwordspacing
G.~Karypis and V.~Kumar, \emph{METIS: A Software Package for Partitioning Unstructured Graphs, Partitioning Meshes, and Computing Fill-Reducing Orderings of Sparse Matrices}, University of Minnesota, Minneapolis, MN, 1998, version 5.1.0. [Online]. Available: \url{http://glaros.dtc.umn.edu/gkhome/metis/metis/overview}
\BIBentrySTDinterwordspacing

\bibitem{cichocki2016tensor}
A.~Cichocki, R.~Zdunek, A.~H. Phan, and S.-i. Amari, ``Tensor networks for dimensionality reduction and large-scale optimization: Part 1 low-rank tensor decompositions,'' \emph{Foundations and Trends® in Machine Learning}, vol.~9, no. 4-5, pp. 249--429, 2016.

\bibitem{hackbusch2015hierarchical}
W.~Hackbusch, \emph{Hierarchical Matrices: Algorithms and Analysis}.\hskip 1em plus 0.5em minus 0.4em\relax Springer, 2015.

\bibitem{woodruff2014sketching}
D.~P. Woodruff, ``Sketching as a tool for numerical linear algebra,'' \emph{Foundations and Trends® in Theoretical Computer Science}, vol.~10, no. 1--2, pp. 1--157, 2014.

\bibitem{cuSPARSE}
\BIBentryALTinterwordspacing
{NVIDIA Corporation}, \emph{cuSPARSE Library}, 2024, version 12.6. [Online]. Available: \url{https://docs.nvidia.com/cuda/cusparse/index.html}
\BIBentrySTDinterwordspacing

\bibitem{magma}
\BIBentryALTinterwordspacing
U.~o.~T. Innovative Computing~Laboratory, \emph{MAGMA Library}, 2024, version 2.8.0. [Online]. Available: \url{https://icl.utk.edu/magma/}
\BIBentrySTDinterwordspacing

\bibitem{NEURIPS2019_9015}
\BIBentryALTinterwordspacing
A.~Paszke, S.~Gross, F.~Massa, A.~Lerer, J.~Bradbury, G.~Chanan, T.~Killeen, Z.~Lin, N.~Gimelshein, L.~Antiga, A.~Desmaison, A.~Kopf, E.~Yang, Z.~DeVito, M.~Raison, A.~Tejani, S.~Chilamkurthy, B.~Steiner, L.~Fang, J.~Bai, and S.~Chintala, ``Pytorch: An imperative style, high-performance deep learning library,'' in \emph{Advances in Neural Information Processing Systems 32}.\hskip 1em plus 0.5em minus 0.4em\relax Curran Associates, Inc., 2019, pp. 8024--8035. [Online]. Available: \url{http://papers.neurips.cc/paper/9015-pytorch-an-imperative-style-high-performance-deep-learning-library.pdf}
\BIBentrySTDinterwordspacing

\bibitem{jax2018github}
\BIBentryALTinterwordspacing
J.~Bradbury, R.~Frostig, P.~Hawkins, M.~J. Johnson, C.~Leary, D.~Maclaurin, G.~Necula, A.~Paszke, J.~VanderPlas, S.~Wanderman-Milne, and Q.~Zhang, ``Jax: composable transformations of python+numpy programs,'' 2018. [Online]. Available: \url{http://github.com/google/jax}
\BIBentrySTDinterwordspacing

\end{thebibliography}

\end{document}